\documentclass[runningheads]{llncs}
\usepackage[T1]{fontenc}
\usepackage{graphicx}
\usepackage{makecell}
\usepackage{booktabs}
\usepackage{xcolor}
\usepackage{subfigure}
\usepackage{hyperref}
\usepackage[smartEllipses]{markdown}
\begin{document}
\title{LLMs as Educational Analysts: Transforming Multimodal Data Traces into Actionable Reading Assessment Reports}

\titlerunning{Transforming Multimodal Data into Actionable Reports using LLMs}

\newif\ifanon

%
%
\ifanon
\else
    \author{Eduardo Davalos\inst{1}\orcidID{0000-0001-7190-7273} \and
    Yike Zhang\inst{1}\orcidID{0000-0003-3503-2996} \and
    Namrata Srivastava\inst{1}\orcidID{0000-0003-4194-318X} \and
    Jorge Alberto Salas\inst{1}\orcidID{0000-0002-8885-0813} \and
    Sara McFadden\inst{1}\orcidID{0000-0001-6375-1272} \and
    Sun-Joo Cho\inst{1}\orcidID{0000-0002-2600-8305} \and
    Gautam Biswas\inst{1}\orcidID{0000-0002-2752-3878} \and
    Amanda Goodwin\inst{1}\orcidID{0000-0002-6439-7399}}
    
    \authorrunning{Davalos et al.}

    \institute{Vanderbilt University, Nashville TN 37235, USA \and
    \email{\{eduardo.davalos.anaya, yike.zhang, namrata.srivastava, jorge.a.salas, sara.mcfadden, sj.cho, gautam.biswas, amanda.goodwin\}@vanderbilt.edu}\\
    \url{http://www.vanderbilt.edu}}

    \newcommand{\school}{Vanderbilt University}

    \newcommand{\githubrepo}{\href{https://github.com/edavalosanaya/LLMsAsEducationalAnalysts}{https://github.com/edavalosanaya/LLMsAsEducationalAnalysts}}
\fi

\maketitle              
\begin{abstract}
Reading assessments are essential for enhancing students' comprehension, yet many EdTech applications focus mainly on outcome-based metrics, providing limited insights into student behavior and cognition. This study investigates the use of multimodal data sources -- including eye-tracking data, learning outcomes, assessment content, and teaching standards -- to derive meaningful reading insights. We employ unsupervised learning techniques to identify distinct reading behavior patterns, and then a large language model (LLM) synthesizes the derived information into actionable reports for educators, streamlining the interpretation process. LLM experts and human educators evaluate these reports for clarity, accuracy, relevance, and pedagogical usefulness. Our findings indicate that LLMs can effectively function as educational analysts, turning diverse data into teacher-friendly insights that are well-received by educators. While promising for automating insight generation, human oversight remains crucial to ensure reliability and fairness. This research advances human-centered AI in education, connecting data-driven analytics with practical classroom applications.

\keywords{LLM, multimodal learning analytics, multi-agents, eye-tracking, assessment, report, unsupervised clustering}
\end{abstract}

\section{Introduction}
Assessing reading proficiency is vital for tracking student progress and informing instructional strategies. Traditional assessments often rely on unimodal metrics like test scores and completion times, which overlook the complex cognitive and behavioral processes involved in learning \cite{Blikstein2013,Worsley2012}. This limitation results in insufficient insights for teachers, making it hard to identify specific student challenges in comprehension, decoding, or fluency. By incorporating multimodal data and analytics, we can gain a more comprehensive understanding of students' learning processes and provide better guidance for instruction \cite{Ochoa_Worsley_2016,cohn2024multimodalmethodsanalyzinglearning}.

Eye-tracking technology is a promising tool for non-intrusively analyzing reading behaviors and cognitive processes  \cite{Just1980AComprehension,Srivastava2018CombiningRecognition}. Metrics such as dwell time, text coverage, and words per minute (WPM) offer insights into how students interact with text, revealing patterns that can indicate comprehension, attention, and decoding difficulties \cite{Southwell2020WhatTexts,hutt_automated_2019,ozeri-rotstain_relationship_2020}. However, the complexity of analyzing raw eye-tracking data can make it difficult for teachers to interpret and apply these insights in the classroom \cite{davalos2025instructedtasksrecognizinginthewild}.

\begin{figure}
    \centering
    \includegraphics[width=\linewidth]{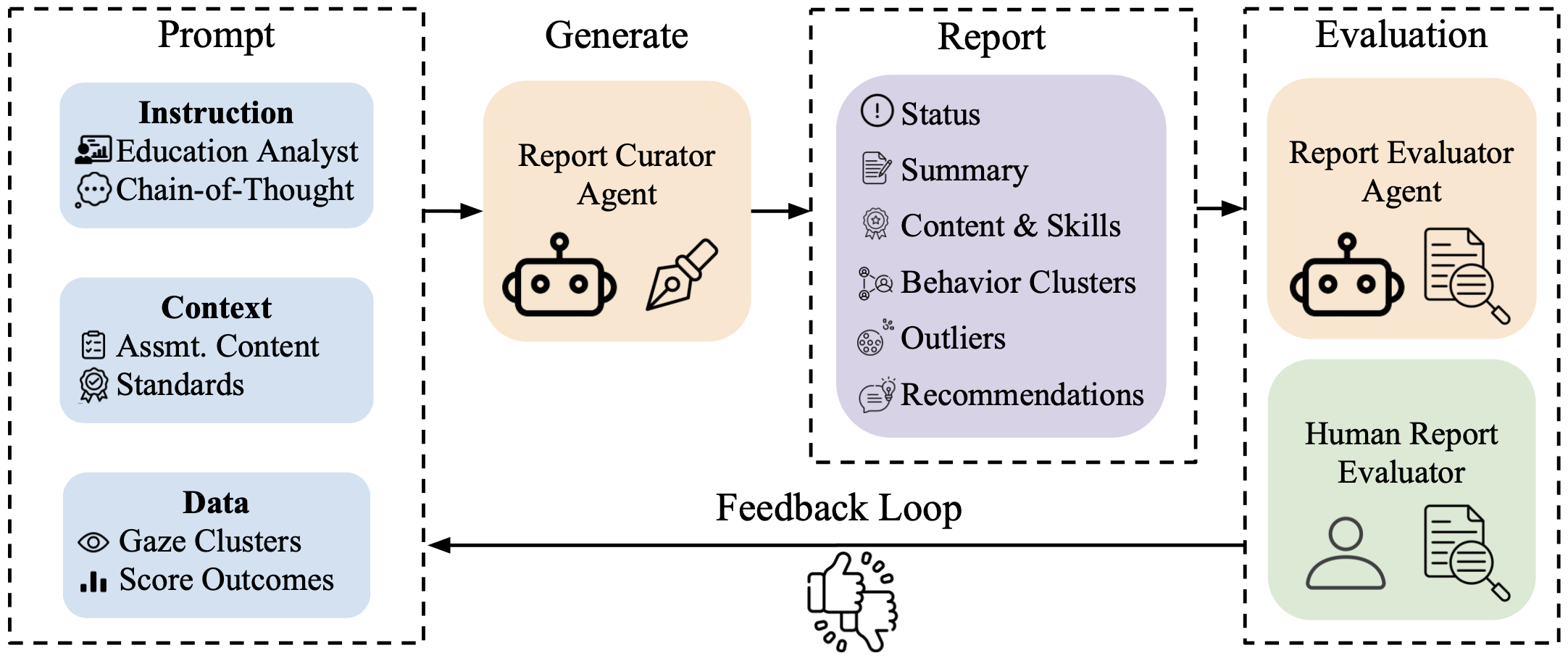}
    \caption{\textbf{Proposed Pipeline for LLM-Driven Assessment Report Generation}: By instructing LLMs to role-play as an educational analyst and providing assessment context and data, we construct a prompt that is used to generate a teacher-oriented assessment report.}
    \label{fig:proposed_pipeline}
\end{figure}

To bridge this gap, we propose a human-in-the-loop AI approach that uses LLMs to generate clear post-assessment reports using eye-tracking data, student learning outcomes, and teaching standards. This can help teachers identify students' reading challenges by turning complex metrics into structured, useful insights. An overview of the proposed system is shown in Fig. \ref{fig:proposed_pipeline}.

This study explores how these AI-generated reports can support teachers in making data-driven instructional decisions while ensuring human oversight for safety and reliability. The central research questions we investigate are:

\begin{itemize}
    \item \textbf{RQ1}: How can unsupervised clustering of eye-tracking features and LLM-based interpretation create meaningful and interpretable student reading profiles?
    \item \textbf{RQ2}: Can an LLM effectively translate classroom reading analytics into actionable teacher reports?
    \item \textbf{RQ3}: How can human-guided and LLM-based evaluations be combined to enhance the clarity, relevance, and usefulness of LLM-generated reports?
\end{itemize}

Integrating process-based assessment with AI-powered reporting, this research advances human-centered AI in education, helping teachers make personalized instructional decisions while reducing the burden of data analysis \cite{Yan2024}. Our source code, template prompts, surveys, teacher responses, and datasets have been made publicly available at 
GitHub\footnote{\githubrepo}.

\section{Related Work}
\subsection{Eye-Tracking and Multimodal Learning Analytics}
Eye-tracking is widely used in learning analytics to identify reading behaviors, cognitive load, and comprehension patterns \cite{Chen2023CharacteristicsStudy,Meziere2023UsingComprehension,Meziere2024ScanpathComprehension}. By capturing real-time gaze data, researchers analyze how students process information, allocate attention, and navigate content. In reading assessment, eye-tracking reveals insights into reading fluency, comprehension difficulties, and attention regulation through metrics like fixation counts, dispersion, and saccadic movement \cite{Southwell2020WhatTexts}. These metrics have been used to distinguish high and low-performing readers \cite{jian_eye-movement_2017} and assess cognitive effort in complex tasks \cite{Chen2021}.

Machine learning (ML) techniques have been used to analyze eye-tracking data and predict learning variables, such as comprehension scores \cite{Meziere2024ScanpathComprehension}, reading behavior \cite{Davalos2023IdentifyingGraphs}, and mind-wandering \cite{hutt_automated_2019}. Fully supervised ML methods are powerful in directly regressing learning variables, but they require laborious human annotations and are prone to overfitting on a specific dataset. Unsupervised clustering does not need labels and has been explored to identify student behavior profiles \cite{peach_data-driven_2019}, but these methods need manual interpretation, limiting scalability in automated educational systems \cite{Davalos2023IdentifyingGraphs}. Our work combines unsupervised clustering with LLMs to automatically interpret and describe clusters, reducing human annotation and making data-driven insights more accessible to educators.

A key limitation in existing eye-tracking research is its lack of direct applicability for teachers. Eye-tracking provides rich data, but current approaches prioritize statistical and data-driven analyses of underlying behaviors over generating teacher-friendly insights \cite{Meziere2024ScanpathComprehension}. Without structured, interpretable feedback, teachers struggle to translate gaze metrics into meaningful instructional interventions \cite{Southwell2020WhatTexts,knoop-van_campen_how_2021}. This gap underscores the need for systems that process eye-tracking data to support teacher decision-making.

\subsection{Generative AI (GenAI) and Education}
Recent advances in GenAI have introduced applications in education, such as automated feedback generation, student support systems, and data-driven instructional assistance \cite{HWANG2020,yan_promises_2024}. AI tools now generate personalized course planning \cite{Hu2024}, grade student responses \cite{Cohn_Hutchins_Le_Biswas_2024}, and provide automated tutoring \cite{molina2024leveragingllmtutoringsystems}, reducing teachers' manual workload. AI-driven solutions, like chatbots for data visualization literacy and dashboard analysis, support teachers in interpreting complex data. VizChat, for example, has shown how conversational AI systems help educators interact with performance dashboards and understand visual learning analytics \cite{Yan2024}. 


Despite GenAI's rise in education, gaps remain. The structuring and interpretation of eye-tracking features for LLMs are unexplored, limiting gaze data used in AI reporting. Multimodal AI approaches for classroom-wide reading assessment reports have yet to be studied, despite LLMs’ success in text summarization. Challenges like hallucinations, bias, and explainability pose risks when AI insights inform instruction, necessitating a human-in-the-loop approach for accuracy and interpretability \cite{yan_promises_2024,Fonteles2024}.
This study explores how LLMs can process eye-tracking data to generate pedagogically useful reports, ensuring AI-assisted reading assessment remains interpretable, actionable, and teacher-friendly.

\section{Methodology}


\subsection{Study \& Participants}
This study involved fifth-grade students from a southeastern U.S. school. The \school's Institutional Review Board (IRB) approved the study for ethical compliance. Participants performed three different assessment: ``GatesS'' ($N=36$), ``GatesT'' ($N=36$), and ``Suffrage'' ($N=46$) assessments. The GatesS and GatesT reading assessments each involved reading three one-page passages from the Gates MacGinitie Standardized reading assessments \cite{GatesMacGinitie1969}. Suffrage involved reading a three-page passage on Women’s Suffrage adapted from the National Assessment of Educational Progress (NAEP) \footnote{\href{https://nces.ed.gov/nationsreportcard/}{nces.ed.gov/nationsreportcard/}}. The aim of these silent reading assessments is to evaluate reading comprehension and behavioral engagement.  

Students participated in two phases of reading tasks. In the cold read phase, they read a passage without viewing comprehension questions, allowing researchers to observe natural reading behaviors. After reading, they entered the question-answering phase, responding to multiple-choice questions related to the text chronologically. Students could revisit the text while answering questions but were unable to return to previous questions, preventing them from using later questions to inform their earlier responses.  

Data was collected using RedForest \cite{davalos_gazeviz_2024}, a computer-based learning environment that managed the assessments and logged interactions. Eye-tracking data were recorded using the Tobii Pro Spark, capturing visual attention patterns. The study used HP 15-dy2795wm laptops with screen dimensions of 19.5 × 34.5 cm and a 704 × 1528 pixels resolution. Additionally, reading scores, item responses, and action logs detailing student activities, including reading time and response latencies, were recorded. 

\subsection{Data Preprocessing}

\begin{figure}
    \centering
    \includegraphics[width=\linewidth]{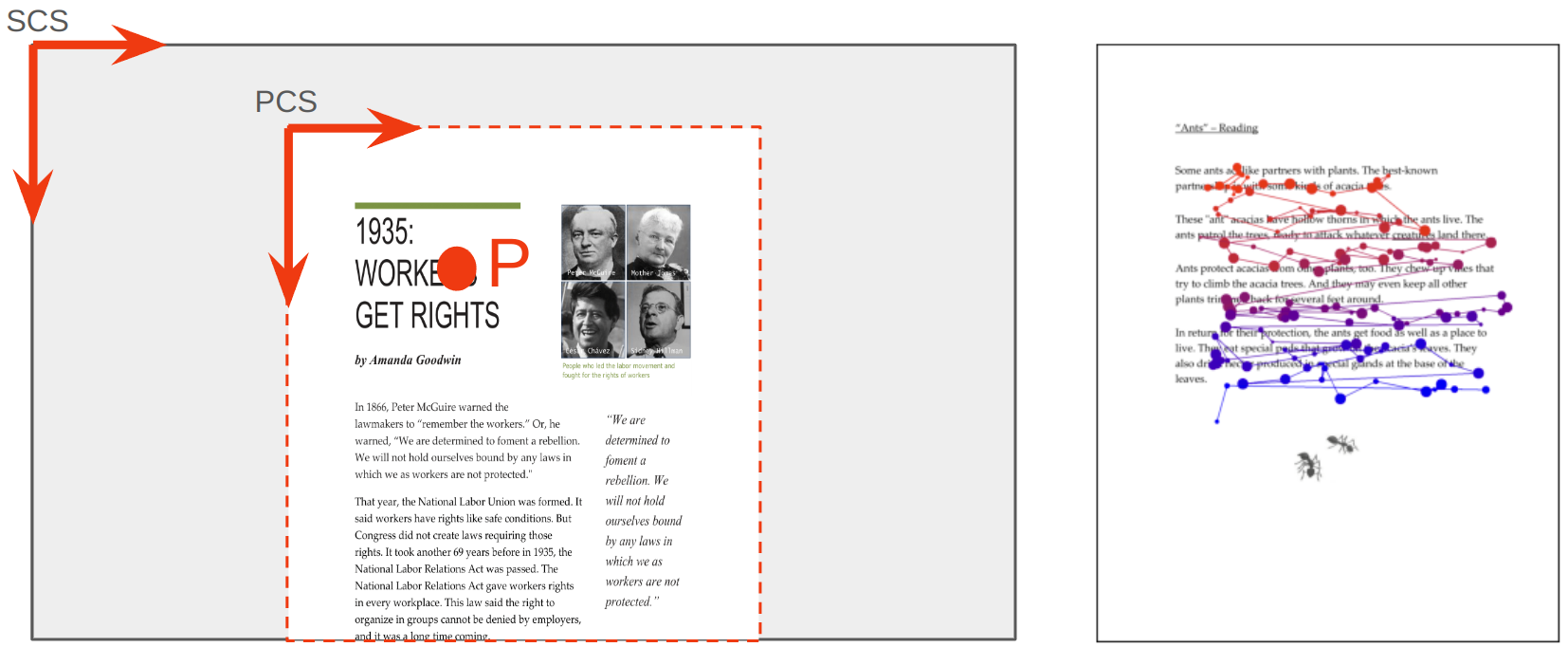}
    \caption{\textbf{Real-Time AOI Encoding}: Performed in-the-moment preprocessing by assigning passage or quiz AOI for each incoming gaze point.
}
    \label{fig:preprocessing_aoi}
\end{figure}

The eye-tracking data from the Tobii Pro Spark system was initially recorded in screen coordinates, mapping gaze points to the laptop display. Fixation detection was performed using the open-source Tobii I-VT algorithm \cite{olsen_tobii_2012}, optimized for screen-based eye trackers, with default parameter settings. The computed fixations were aligned with predefined Areas of Interest (AOIs) for the passage text and quiz interface to enable analysis. This encoding linked each fixation to a specific region, differentiating reading behaviors across task segments. The mapping process enhanced the tracking of student interactions with the passage content and comprehension questions, as shown in Fig. \ref{fig:preprocessing_aoi}. 

After AOI encoding, gaze data were aggregated to create metrics describing students' reading behaviors during the cold read and question-answering phases. Key indicators included fixation dispersion, saccade regression rate, text coverage, and words per minute (see Table \ref{tab:gaze_features}). Segmenting these features by task phase revealed distinct patterns related to reading comprehension and response processing.

\begin{table}[]
\caption{\textbf{Gaze-Based Extracted Features}: After preprocessing, we compute gaze-based features linked to reading processes. These accumulative metrics are generated for the cold read or question-answering segments of the reading assessment.}
\centering
\begin{tabular}{p{0.30\linewidth} p{0.65\linewidth}}
\toprule
Variable                & Description \\
\midrule
Gaze-based WPM          & WPM is computed in 10-second windows by dividing AOI travel distance by fixation duration, with the median across windows serving as the final metric. \\
Line Coverage           & Fraction of line AOI the participant fixated on \\
Dwell Time              & Summation of fixation durations for all fixations that fall within an AOI (quiz/passage) \\
Fixation Dispersion     & Fixation dispersion is computed in 10-second windows as the average Euclidean distance from the centroid of all fixations within each window. \\
Saccade Regression Rate & Percentage of backward saccade of any length \\
\bottomrule
\end{tabular}
\label{tab:gaze_features}
\end{table}

\subsection{Student Grouping via Clustering} 
\begin{figure}
    \centering
    \includegraphics[width=\linewidth]{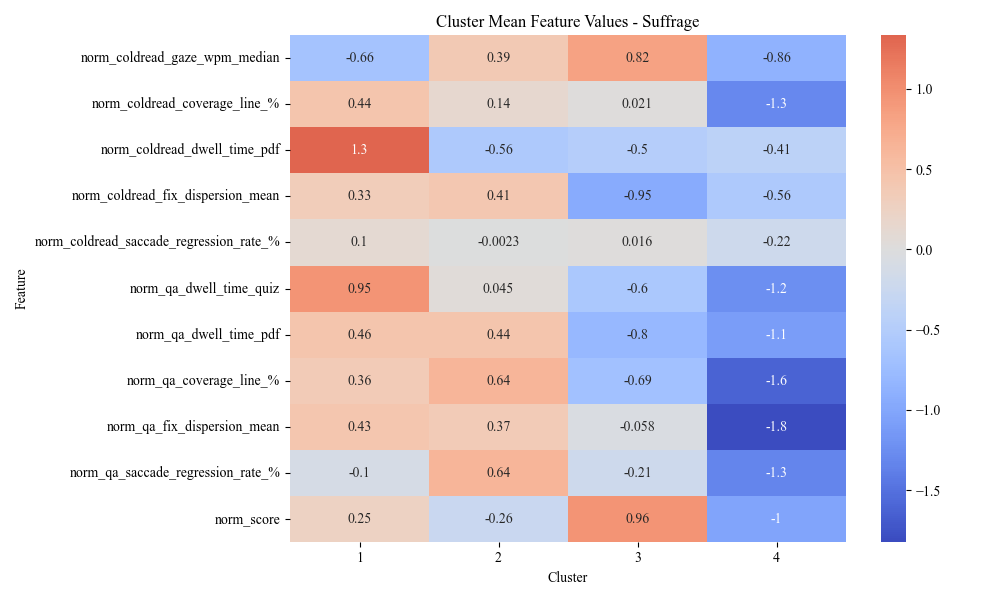}
    \caption{\textbf{K-Means Example Cluster Feature Heatmap}: Scaled gaze features are input into an unsupervised clustering algorithm. The cluster centroids are used to construct profiles based on gaze features, resulting in a reading behavior profile for each cluster.
}
    \label{fig:cluster_heatmap_example}
\end{figure}

An unsupervised learning approach is applied to the eye-tracking features in Table \ref{tab:gaze_features} to identify patterns in student reading behavior. We analyze standardized values of these features to gain insights into different reading strategies. An example heat map corresponding to the clusters generated is shown in Fig. \ref{fig:cluster_heatmap_example}. Instead of manual labeling, we use an LLM to interpret each cluster’s behavioral profile, assigning descriptors like ``Careful Reader'' or ``Rapid Scanner'' based on eye-tracking patterns. This enhances the explainability of results for educators.

To determine the most effective clustering technique for identifying reading behavior profiles, we evaluated three similar unsupervised algorithms: K-Means, Gaussian Mixture Models (GMM), and Spectral Clustering. We based our selection on clustering quality metrics, including within-cluster variance \cite{grira2004unsupervised} for compactness and silhouette scores \cite{ROUSSEEUW198753} for cluster separation.

\subsection{LLM Agents: Report Curator and Evaluator}
 
The LLM prompting system underwent iterative refinement through rapid prompt engineering and continuous educator feedback. By incorporating teacher insights at each stage, the final framework generates actionable reports linking student performance to education standards.  

The system employs two specialized LLM agents: a \textbf{Report Curator} and a \textbf{Report Evaluator}. The Report Curator generates student performance reports by synthesizing multiple input sources, structured in JSON format, to produce a comprehensive analysis of classroom and individual student performance. Inputs include teaching principles like Common Core Reading Standards\footnote{\href{https://www.thecorestandards.org/ELA-Literacy/}{www.thecorestandards.org/ELA-Literacy/}} and grade-level reading fluency skills\footnote{\href{https://www.tn.gov/education/districts/academic-standards/english-language-arts-standards.html}{www.tn.gov/education/districts/academic-standards/english-language-arts-standards.html}}, text complexity metrics, score distributions, question-level performance, and clustering insights. 

The Report Curator's output follows a structured, multi-layered Markdown format to enhance interpretability for educators. The report includes a chain-of-thought reasoning process \cite{Cohn_Hutchins_Le_Biswas_2024}, ensuring transparency in conclusions. It covers students' strengths and weaknesses, content area influences, skill assessments, cluster overviews, and outlier identification, with recommendations for targeted support. The report concludes with overall instructional recommendations for teachers.

The Report Evaluator assesses the quality of the generated reports using the original prompt and predefined rating criteria, including clarity, relevance, coherence, applicability, depth of insight, specificity, engagement, bias and fairness, and use of evidence. The evaluator produces a structured JSON output with scores and justifications, providing immediate feedback for refining report language and structure. This iterative process addressed issues like variable definitions and layout readability.

Reports were generated using OpenAI's GPT-4o model \cite{openai2024gpt4technicalreport}, with the JSON prompt converted into a formatted string and sent to OpenAI's API \footnote{\href{https://platform.openai.com/docs/overview}{platform.openai.com/docs/overview}}. The GPT-4o's response was parsed to extract the Markdown-formatted report, which was then rendered into a PDF for presentation to teachers.

\subsection{Teacher Survey}

To evaluate the usefulness and interpretability of the reports generated by the LLM, we conducted a survey among teachers centered around the ``Suffrage” classroom assessment report, shown in Fig. \ref{fig:markdown_report}. They assessed six sections: (1) keyword assessment status, (2) summary, (3) content and skill analysis, (4) clusters, (5) outliers, and (6) recommendations, rating the helpfulness on a five-point Likert scale and providing additional comments.

The survey also included broader qualitative questions to gather overall impressions. Teachers identified specific insights that were valuable or difficult to interpret and suggested improvements. The survey concluded with an open-ended prompt for additional feedback.

\begin{figure}
    \centering
    \begin{minipage}{0.49\textwidth} 
        \scriptsize
        \input{reports/Suffrage.tex}
    \end{minipage}
    \hfill
    \begin{minipage}{0.49\textwidth} 
        \scriptsize
        \input{reports/Suffrage2.tex}
    \end{minipage}
    \caption{\textbf{LLM-Generated Suffrage Assessment Report}}
    \label{fig:markdown_report}
\end{figure}

\section{Results}

\subsection{Evaluation of Unsupervised Clustering}

The results of the cluster quality evaluation are summarized in Table \ref{tab:cluster_quality_metrics}, comparing the performance of different methods. K-Means was selected as the best clustering technique due to its highest silhouette score and moderate within-cluster variance, effectively balancing separation and cohesion. Unlike Spectral Clustering and Gaussian Mixture Models (GMM), K-Means generated fewer clusters, avoiding prompt expansion issues that led to hallucinations in the Report Curator Agent. One-way ANOVA tests \cite{Ross2017} showed that 8 out of 10 input features were statistically significant ($p \leq 0.05$) across clusters, as detailed in Table \ref{tab:kmeans_stats}. Overall, K-Means emerges as the most effective choice, offering a balance of interpretability, efficiency, and low cluster counts.

\begin{table}[]
\caption{\textbf{Cluster Quality Metrics}: The cluster count, average within-cluster feature variance, and silhouette scores were computed for K-Means, Gaussian Mixture Model (GMM), and Spectral algorithms. These metrics are for the \textbf{Suffrage} assessment, similar performance \& trends match for the other two assessments.}
\centering
\begin{tabular}{l|c|c|c}
\toprule
Method     & \makecell{Number of \\ Clusters} & \makecell{Avg Within-Cluster \\ Variance} & \makecell{Silhouette \\ Score} \\
\midrule
K-Means     & 4                                      & 0.67                                            & 0.16                                 \\
GMM        & 8                                      & 0.5                                             & 0.13                                 \\
Spectral   & 6                                      & 0.68                                            & 0.16                                 \\
\bottomrule
\end{tabular}
\label{tab:cluster_quality_metrics}
\end{table}

\begin{table}[]
\caption{\textbf{K-Means Statistical Testing}: To show between-cluster differences, we used one-way ANOVA tests across input features between all clusters. For K-Means, all but two features are statistically significant ($p \leq 0.05$).}
\centering
\begin{tabular}{lrr}
\toprule
Feature                                       & \multicolumn{1}{l}{F statistic} & \multicolumn{1}{l}{p-value} \\
\midrule
norm\_qa\_coverage\_line\_\%                  & 19.58                           & \textless{}0.001            \\
norm\_coldread\_coverage\_line\_\%            & 17.26                           & \textless{}0.001            \\
norm\_qa\_saccade\_regression\_rate\_\%       & 11.53                           & \textless{}0.001            \\
norm\_qa\_fix\_dispersion\_mean               & 9.34                            & \textless{}0.001            \\
norm\_qa\_dwell\_time\_pdf                    & 6.45                            & 0.002                       \\
norm\_qa\_dwell\_time\_quiz                   & 4.72                            & 0.008                       \\
norm\_coldread\_gaze\_wpm\_median             & 3.98                            & 0.016                       \\
norm\_coldread\_saccade\_regression\_rate\_\% & 3.36                            & 0.031                       \\
norm\_coldread\_fix\_dispersion\_mean         & 2.85                            & 0.053                       \\
norm\_coldread\_dwell\_time\_pdf              & 1.02                            & 0.396                       \\ 
\bottomrule
\end{tabular}
\label{tab:kmeans_stats}
\end{table}

\subsection{LLM and Human Evaluation of Assessment Reports}

 
To assess the quality of the generated reports, we combined human expertise with the capabilities of a Report Evaluator Agent, leveraging educators' pedagogical insights and the LLM's rapid, scalable evaluations. The Report Evaluator initially evaluated the assessment reports' clarity, coherence, and instructional relevance. As we refined our prompts based on educator feedback, the Report Evaluator’s ratings improved, shifting from lower to higher values on the Likert scale, indicating more pedagogically aligned and interpretable reports over time.  

\begin{figure}
\centering
\subfigure[LLM Global Evaluation on Suffrage, GatesS, and GatesT assessment reports.]{\label{fig:llm_evaluation}\includegraphics[width=60mm]{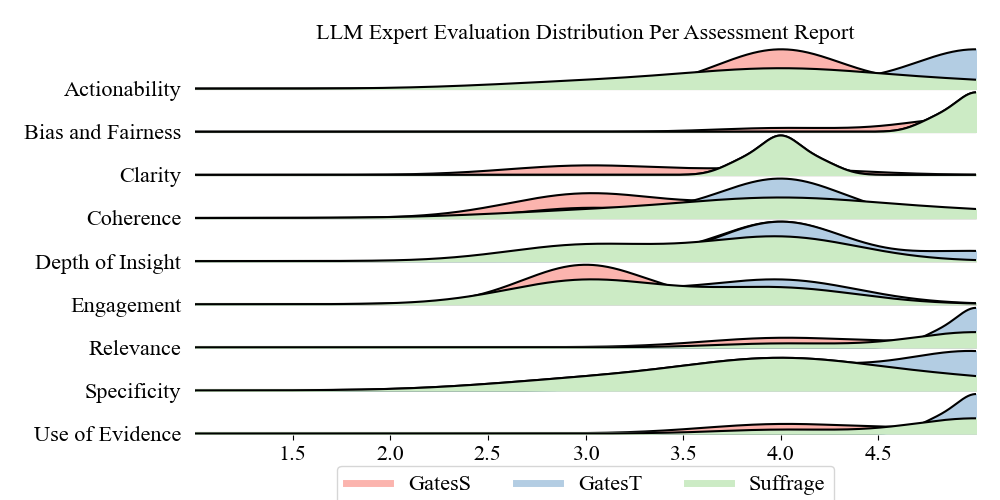}}
\subfigure[Human Per-Section Evaluation on Suffrage assessment report.]{\label{fig:human_evaluation}\includegraphics[width=60mm]{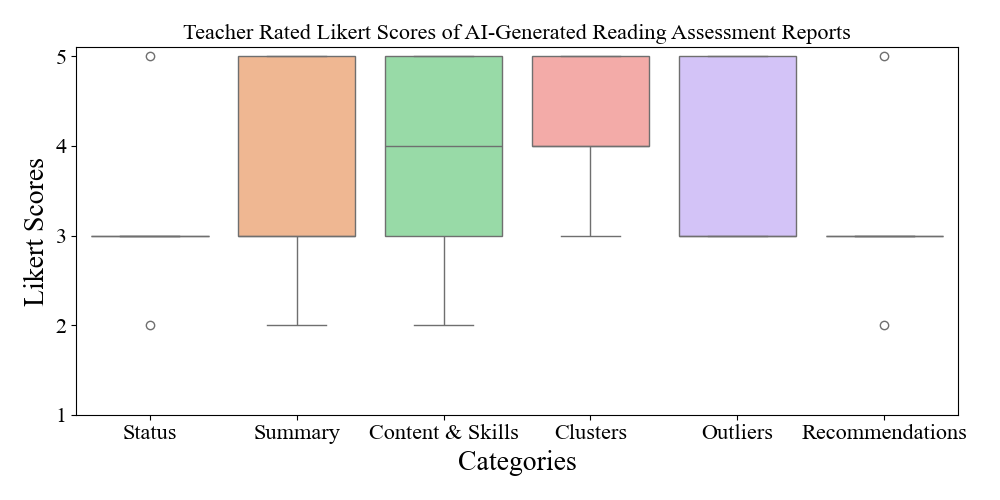}}
\caption{\textbf{LLM and Teacher Evaluations of the Assessment Reports}: A complementary evaluation approach, where the LLM provides an global evaluation of report quality, while teachers offer a detailed, section-by-section analysis of its usefulness and clarity.}
\end{figure}

For consistency, the Report Evaluator was run five times per assessment, generating multiple perspectives on each report's quality. The distribution of evaluations, shown in Figure \ref{fig:llm_evaluation}, illustrates overall rating trends. On average, the Report Evaluator assigned a Likert score of 4, indicating that the reports were generally clear, relevant, and useful for instructional support.

Teacher feedback (N=5, E1-5) provided valuable insights into the strengths and areas for improvement in the LLM-generated assessment reports. Overall, teachers found the \textbf{clusters ($\mu = 4.20\pm0.83$)} and \textbf{outliers section ($\mu = 3.80\pm1.09$)} particularly insightful, highlighting their effectiveness in identifying student reading patterns and those needing additional support. Teachers appreciated the intuitions generated by the clusters, though they suggested minor refinements to align terminology with instructional language. Educators found the information clear and actionable for outliers but recommended emphasizing follow-up assessments rather than drastic interventions.

Teachers appreciated the \textbf{content and skills section ($\mu = 3.80\pm1.30$)}, with teachers noting that the inclusion of reading standards made the report particularly pedagogically relevant and actionable. However, they recommended improving formatting, such as presenting standards in a structured list and providing hyperlinks to specific learning objectives, questions, and data evidence. The \textbf{summary section ($\mu = 3.60\pm1.34$)} contained valuable information, but teachers found the formatting dense, describing it as a ``wall of text.'' They suggested using bullet points, bolding, and simplified language to improve readability.

The \textbf{status section ($\mu = 3.20\pm1.09$)} was seen as a helpful overview. However, teachers suggested refining status labels to be more precise (e.g., replacing ``good comprehension'' with ``grade-level comprehension'') and including percentage-based indicators to better understand student performance. Similarly, the \textbf{recommendations section ($\mu = 3.20\pm1.09$)} was viewed as informative but too generic, with teachers proposing a table format for better clarity.

Teachers provided positive overall impressions of the LLM-generated reports while identifying areas for improvement. One educator (E4) remarked, \textit{``I think the way you've structured this is thoughtfully designed and could definitely be helpful. Generally, I think making the language even more direct will make it both more helpful and user-friendly for teachers to engage with.''} This feedback highlights the importance of clear and direct language to ensure the reports are accessible and actionable for educators.  

Regarding report readability, teachers noted that the formatting could be improved to enhance usability. E3 suggested, \textit{As a whole, the report would benefit from different formatting. The wall of text is overwhelming and might not feel approachable for educators. Creating new paragraphs, bullet points, bolding, etc., to differentiate the text would really help.''} This aligns with prior feedback emphasizing structural refinements to make key insights more digestible.  

Teachers generally found the reports to be valuable in understanding student performance, with E3 stating, \textit{``This provides very helpful feedback on how students performed on this assessment.''} Among the report sections, clustering was the most well-received, with four out of five teachers identifying it as the most useful component. E5 specifically noted, \textit{``I liked the clusters section because it gave a succinct overview with identified students.''} This suggests that providing meaningful student profiles enhances instructional planning by helping teachers quickly categorize and support students based on their reading behaviors.  

An important opportunity for improvement emerged regarding the practical implementation of recommendations. While teachers found the recommendations valuable, they noted that acting on them would require additional training and time, potentially increasing their workload. E3 expressed concern, stating, \textit{``As a teacher, I don't know how much I would use the recommendations. Particularly because 1) how will I be trained on them and 2) now I have to investigate on my own time what these recommendations mean.''} Future iterations should integrate system-assisted recommendations, enabling on-demand support and automated interventions, ensuring they are both actionable and immediately applicable while minimizing teachers workload additions.

\section{Discussion}

Our approach in \textbf{RQ1} showed that using the cluster centroid descriptions as behavioral profiles enabled the LLM to generate meaningful interpretations for each cluster. This method produced labels like ``Steady Comprehenders'' and ``Engaged but Inconsistent'', which helped educators quickly identify student patterns and guide support strategies. By simplifying data-driven segmentation, clustering improved the explainability of student performance, making insights actionable for teachers.

For \textbf{RQ2}, our findings suggest that LLMs effectively synthesize multimodal learning analytics into structured, teacher-friendly reports. Educators responded positively, especially to student clustering insights and content analysis, which linked behaviors to instructional needs. While teachers valued access to raw data, LLM-generated summaries streamlined decision-making. However, feedback highlighted the need for refined terminology, improved layout, and recommendations that minimize additional workload.

For \textbf{RQ3}, combining LLMs and human expertise created an efficient evaluation pipeline that improved report quality while reducing teacher workload. LLMs provided rapid feedback on clarity and structure, enabling quick refinements, while teachers ensured pedagogical relevance by aligning reports with instructional needs. This hybrid approach streamlined iterations, balancing automation with expert oversight to produce reports that were both readable and relevant for instructors.


\section{Conclusions and Future Work}

This study introduced a gaze-based clustering approach combined with an LLM-driven educational analyst to interpret student reading behavior profiles and synthesize multimodal data into teacher-friendly reports. By leveraging unsupervised clustering, the system identified meaningful student groups, while the LLM transformed raw analytics into interpretable and actionable insights. Empirical validation through educator evaluations revealed that teachers found these reports accessible, succinct, and helpful, particularly valuing the cluster profiles and content analysis for guiding instructional decisions. These findings highlight the importance of human-AI collaboration in educational analytics, demonstrating how AI-driven insights can complement teacher expertise to enhance classroom decision-making.

These reports should spark an ongoing teacher-LLM dialogue about classroom performance. Future work will enable teacher-driven inquiries for real-time, data-informed responses and integrate on-demand intervention support and a more conversational interface to enable teachers to ask follow-up questions that seek clarifications and more details. The system can provide actionable, teacher-centric decision support by combining LLM-driven insights with structured analytics and automated support without adding to educators' workload.

\subsubsection{\ackname} 
\ifanon
\else
The research reported here was supported by the Institute of Education Sciences, U.S. Department of Education, through Grant R305A150199 and R305A210347 to Vanderbilt University. The opinions expressed are those of the authors and do not represent views of the Institute or the U.S. Department of Education. 
\fi

%
%
%
\bibliographystyle{splncs04}
\bibliography{references}
\end{document}